\documentclass{article}
\usepackage[preprint]{neurips_2026}
\usepackage[T1]{fontenc}
\usepackage[utf8]{inputenc}
\usepackage{lmodern,microtype}
\usepackage{amsmath,amssymb,amsthm,graphicx,booktabs,tabularx,array,multirow}
\usepackage{caption,placeins,xspace,needspace,float}
\usepackage{xcolor,hyperref}
\hypersetup{colorlinks=true,linkcolor=blue!55!black,citecolor=blue!55!black,urlcolor=blue!55!black,
pdftitle={JEV-VAD: Efficient Semantic Endpointing via Task-First Conditioning and Decoder-Prefix Pruning},
pdfauthor={Jintao Liang, Nan Li, Longbiao Wang, Tubin Huang, Min Xu}}
\setcitestyle{authoryear,square}
\newcommand{\method}{\textsc{JEV-VAD}\xspace}
\newcommand{\sg}{\operatorname{sg}}
\newcommand{\ind}{\mathbf{1}}

\title{{\fontsize{15}{18.5}\selectfont
JEV-VAD: Efficient Semantic Endpointing via\\
Task-First Conditioning and Decoder-Prefix Pruning}}
\author{Jintao Liang$^1$ \quad Nan Li$^{2,*}$ \quad Longbiao Wang$^1$\\
\textbf{Tubin Huang}$^3$ \quad \textbf{Min Xu}$^3$\\[5pt]
\small $^1$Tianjin University \quad $^2$Tianjin Normal University\\
\small $^3$Shanghai Shenqijiyuan Intelligent Terminal Co., Ltd.\\
\small $^*$Corresponding author: \texttt{linan@tjnu.edu.cn}}
\date{}
\begin{document}
\raggedbottom
\maketitle
\begin{abstract}
Semantic endpointing must distinguish a completed turn from a temporary pause, yet its binary output does not require the full generation pathway of an audio--language model. We propose JEV-VAD, an ASR-free framework that combines task-first conditioning, a scalar endpoint readout, and decoder-prefix pruning. The question and completion criteria are placed before audio tokens, allowing causal decoder attention to construct audio states conditioned on the endpointing objective. A dedicated scalar head maps the assistant-prefix state directly to a stop probability, removing transcript generation, answer decoding, and vocabulary-sized output projection. To reduce representation cost, decoder-prefix pruning physically retains only the initial decoder blocks. Matched head refitting then identifies which retained depths provide useful endpoint information under a common adaptation budget. The framework is built on Qwen3-ASR-0.6B and adapted through supervised fine-tuning followed by reinforcement learning for calibrated decisions. Together, these components address input conditioning, output complexity, and decoder depth within one endpointing framework. Experiments on EOT-bench show cross-language transfer from English endpoint supervision and useful discrimination in shallow prefixes, supporting a selectable quality--resource trade-off.
\end{abstract}

\section{Introduction}
\label{sec:intro}
Voice agents must decide when a user has completed a turn before taking the conversational floor. Responding during a hesitation interrupts an unfinished thought; waiting after completion adds unnecessary silence. Human turn-taking draws on both linguistic and temporal information \citep{stivers,levinson,skantze}. A pause therefore creates an opportunity to respond, but does not establish that the speaker has finished. Semantic endpointing addresses this ambiguity by predicting completion from the available speech rather than treating every sufficiently long silence as a turn boundary.

Existing approaches use acoustic timing, lexical information, or learned speech representations. Finite-state and continuous prediction models incorporate timing and prosody \citep{schlangen,raux,skantze2017,maier,roddy}, while language-based approaches model linguistic completion \citep{turngpt}. These signals answer different questions: timing describes how long the user has paused, whereas linguistic context helps determine whether the preceding utterance expresses a complete thought. Transcript-based systems introduce an intermediate recognition output before the endpoint decision. Direct audio endpoint detectors instead seek the relevant evidence in speech itself, as explored by recent semantic detectors and dialogue-state predictors \citep{smartturn,turnsense,phoenix,soulxpaper}.

Pretrained speech models provide a foundation for this direct decision path. Self-supervised representations \citep{wav2vec,hubert,wavlm} and recognition-based representations \citep{whisper,qwen3asr} can be reused for downstream speech understanding. Audio--language systems connect such representations to language processing \citep{salmonn,qwen2audio,moshi,llamaomni,llamaomni2}. However, a semantic endpoint controller requires one completion probability, while the inherited model may contain a full language decoder and vocabulary projection. Eliminating transcript generation simplifies the output path, but leaves a second question: how much decoder computation is necessary to construct the decision state?

Typed audio decisions provide a useful starting point. Qwen-Audio-Jev \citep{qwenjev} introduces task-conditioned readouts, and DuplexJev \citep{duplexjev} obtains option probabilities without autoregressive decoding. Building on this direction, we study three complementary design choices. The first is \emph{where to specify the task}: in a causal decoder, placing the question before audio permits audio positions to attend to the instructions. The second is \emph{what to predict}: a scalar stop probability avoids computing a full answer-token distribution. The third is \emph{how much depth to retain}: a fixed endpointing task may remain accessible to a readout attached to a shallow decoder prefix.

We introduce \method, an ASR-free endpoint detector based on Qwen3-ASR-0.6B. Here, ASR-free means that endpoint inference produces no transcript; the model still benefits from a recognition-pretrained backbone. The question and decision criteria precede the audio tokens, and a scalar head reads the assistant-prefix state. The model is adapted with supervised fine-tuning (SFT) followed by reinforcement learning for calibrated decisions (RLCD). After adaptation, decoder-prefix experiments retain the first $d$ blocks, freeze the retained backbone and LoRA weights, and refit a fresh head with a matched budget. This protocol tests the accessibility of endpoint information at different depths without conflating every prefix with the original final-layer head.

Our experiments separate full-model system comparisons from multi-seed prefix analysis. English-only endpoint supervision is evaluated on 14-language EOT-bench. The full model achieves the lowest Chinese cut@300ms among the compared systems, while the multilingual results also identify stronger proprietary operating points. The prefix experiments quantify the resource reductions and language-dependent changes associated with a seven-block model. Our contributions are:
\begin{enumerate}
\item A task-first, scalar endpointing interface, evaluated against an audio-first input order under matched training conditions.
\item A decoder-prefix study with three backbone seeds and three head seeds per backbone, including a no-attention control and dense depth sampling in the first seven blocks.
\item A multilingual evaluation and explicit parameter/computation accounting that distinguish full-model quality from the bilingual trade-off induced by pruning.
\end{enumerate}

\section{Related Work}
\label{sec:related}
\paragraph{Turn completion and conversational timing.}
Turn-taking research distinguishes the temporal organization of conversation from the semantic and interactional evidence that a turn is complete \citep{stivers,levinson,skantze}. Computational approaches include predictive timing models \citep{schlangen}, finite-state turn management \citep{raux}, and continuous neural prediction using speech features \citep{skantze2017,maier,roddy}. TurnGPT studies linguistic completion using a language model \citep{turngpt}. These lines motivate evaluating endpointing as a trade-off between false interruptions and waiting, rather than as silence classification alone.

\paragraph{Semantic endpoint detectors.}
Smart Turn and TurnSense provide direct audio completion interfaces \citep{smartturn,turnsense}. Phoenix-VAD studies semantic endpointing for full-duplex interaction \citep{phoenix}, while SoulX-Duplug addresses streaming dialogue-state prediction \citep{soulxpaper}. FD-VAD releases an ASR-free endpointing model with a speech encoder and a small language backbone \citep{fdvad}. We evaluate several released systems under EOT-bench, using its published references where local outputs are unavailable. Our focus is the internal computation of a task-conditioned detector, particularly input order, output dimension, and retained decoder depth.

\paragraph{Speech representations and audio--language interfaces.}
wav2vec 2.0, HuBERT, and WavLM learn reusable speech representations \citep{wav2vec,hubert,wavlm}. Whisper and Qwen3-ASR supply recognition-trained representations \citep{whisper,qwen3asr}. SALMONN and Qwen2-Audio connect audio to language-model processing \citep{salmonn,qwen2audio}; Moshi and the LLaMA-Omni family extend speech interaction capabilities \citep{moshi,llamaomni,llamaomni2}. These systems address broader understanding or generation tasks. \method specializes an inherited representation pathway for a single binary control decision.

\paragraph{Direct decisions and reduced decoder depth.}
Qwen-Audio-Jev provides typed audio decision heads and SFT/RLCD objectives \citep{qwenjev}; DuplexJev reads answer-option probabilities without autoregressive generation \citep{duplexjev}. We adopt the scalar-decision perspective for endpointing and examine its interaction with causal input order. LayerSkip studies early-exit inference and self-speculative decoding for language generation \citep{layerskip}. Our setting differs: the pruned model emits one scalar, and each retained prefix receives a separately refitted head. The depth study therefore measures endpoint-readout quality rather than the ability to reproduce full-depth generated text.

\section{JEV-VAD Architecture}
\label{sec:method}

Let $x_t^{(W)}$ be a trailing window of audio available by time $t$. Given a completion question $q$ and criteria $c$, the detector estimates $p_t=f_\theta(x_t^{(W)},q,c)$, with target $y=1$ for a completed turn and $y=0$ for an intermediate pause. A separate policy converts this score into a turn-ending action.
\subsection{Audio pathway and scalar readout}
\begin{figure}[tbp]
\centering
\includegraphics[width=\linewidth]{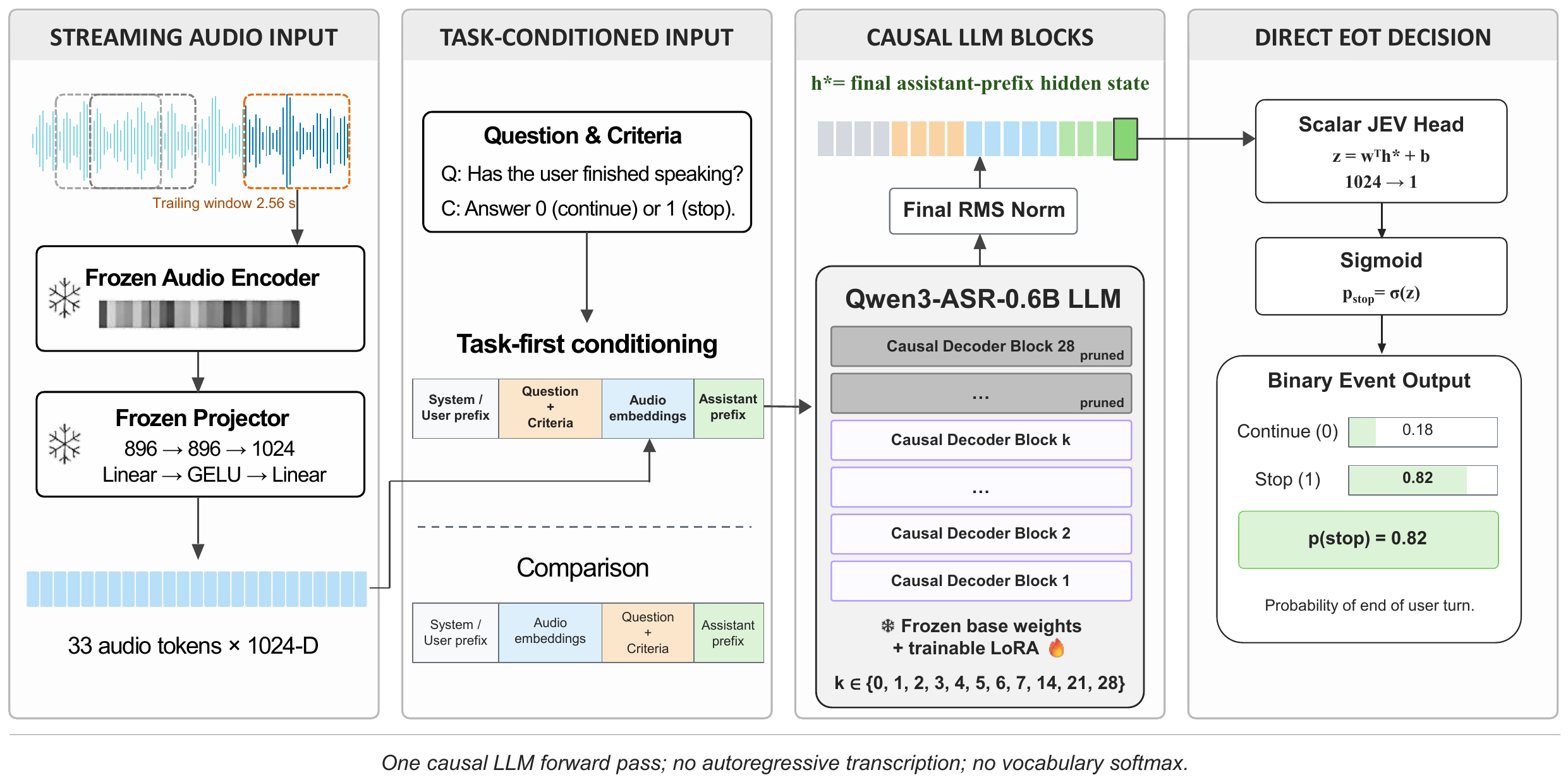}
\caption{\method architecture. The completion question and criteria precede audio tokens. The scalar head reads the assistant-prefix state after RMSNorm. Decoder LoRA is adapted during SFT/RLCD and frozen during prefix-head refitting.}
\label{fig:architecture}
\end{figure}

Figure~\ref{fig:architecture} shows the representation pathway. The base model is Qwen3-ASR-0.6B \citep{qwen3asr}. A frozen audio encoder and the native $896\rightarrow896\rightarrow1024$ GELU projector transform a 2.56 s, 16 kHz window into 33 audio tokens. The decoder receives a system/user prefix, task instructions, audio boundary markers and projected tokens, and an assistant prefix. The final assistant-prefix state is used as the readout position.

Let $h_\star\in\mathbb{R}^{1024}$ denote this state after the final RMSNorm. The scalar endpoint head is
\begin{equation}
z=w^\top h_\star+b,\qquad p_{\mathrm{stop}}=\sigma(z),
\label{eq:head}
\end{equation}
with 1,025 trainable parameters. The head produces a continuous stop probability in one forward pass. It replaces the vocabulary projection and does not generate transcript tokens or an answer string. During endpoint adaptation, only decoder LoRA \citep{lora} and this head are trainable, totaling 4,588,545 parameters. Encoder, projector, and decoder base weights remain frozen.

\subsection{Task-first conditioning}
The question $q$ asks whether speech is finished, and the criteria $c$ map continuation and completion to 0 and 1. In the proposed order, the decoder receives $q+c+\mathrm{audio}$. Its causal attention mask permits each audio position to attend to the earlier question and criteria. In the alternative $\mathrm{audio}+q+c$ order, audio positions cannot attend to the later task instructions, although the final readout can attend to the preceding sequence in both cases.

Both orders use the same token count, readout position, and training procedure. The comparison tests instruction placement; it does not separately isolate semantic and positional effects.

\subsection{Online windowing}
Inference updates every 320 ms using a trailing 2.56 s window. The encoder is bidirectional within the audio that has already arrived, so the overall score remains causal with respect to the input stream even though the encoder is not internally causal. No future audio is supplied to an update.

Short inputs are left-padded to 2.56 s. This leading silence is treated as valid input, whereas padding introduced to form a batch is masked. 

\subsection{Decoder-prefix retention}
Let L$d$ retain the first $d$ of the 28 decoder blocks. We evaluate $d\in\{0,1,2,3,4,5,6,7,14,21,28\}$. Sampling every block below L07 examines early endpoint discrimination; the deeper points cover quarter-depth budgets. Later blocks are physically removed in the pruned model rather than executed and ignored.

Each retained prefix inherits the same two-stage adapted model within its backbone seed. The retained backbone and LoRA weights are frozen, a fresh scalar head is attached after RMSNorm, and the head is refitted with the same update budget at every depth. L28 receives the same refitting treatment. L28 is therefore a matched refitted-head control, distinct from the full-model system-comparison row.

L00 removes all decoder attention. With fixed task tokens and a fixed readout position, that position cannot aggregate audio-token information. L00 therefore provides a no-attention control.

\section{Two-Stage Adaptation}
\label{sec:training}
\begin{figure}[tbp]
\centering
\includegraphics[width=0.95\linewidth]{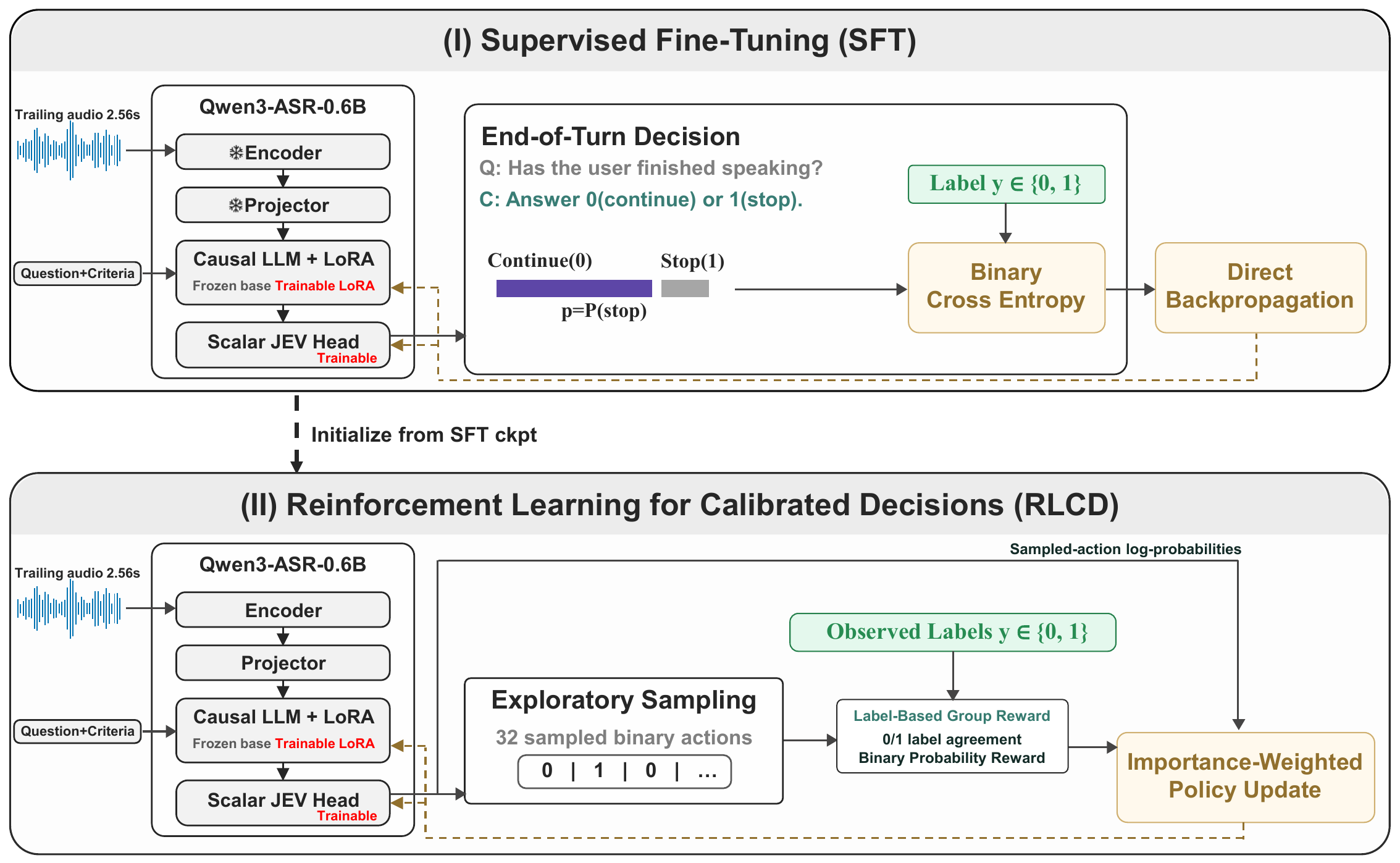}
\caption{Training sequence. Stage I fits the endpoint labels with binary cross-entropy. Stage II initializes from SFT and applies the Bernoulli RLCD objective. All sampled actions share one forward distribution.}
\label{fig:training}
\end{figure}

Figure~\ref{fig:training} summarizes SFT followed by RLCD. The two stages use the same endpoint labels and scalar interface; both update decoder LoRA and the head while leaving the base representation pathway frozen.

\subsection{Stage I: supervised fine-tuning}
Completed turns have $y=1$ and unfinished turns have $y=0$. With $p=p_{\mathrm{stop}}$, Stage I minimizes
\begin{equation}
\mathcal{L}_{\mathrm{SFT}}=-y\log p-(1-y)\log(1-p).
\label{eq:sft}
\end{equation}

\subsection{Stage II: Bernoulli RLCD}
Starting from SFT, we specialize the Noul decision objective of Qwen-Audio-Jev \citep{qwenjev} to
\begin{equation}
\pi(1)=p,\qquad \pi(0)=1-p.
\end{equation}
For each input, $M=32$ actions are sampled from the exploratory behavior policy
\begin{equation}
\mu(a)=(1-\epsilon)\pi(a)+\epsilon/2,\qquad \epsilon=0.05,
\end{equation}
with importance weights $w_r=\pi(a_r)/\mu(a_r)$. Define the weighted agreement terms
\begin{align}
C_{-r}&=\sum_{s\ne r}w_s\ind[a_s=y],\\
P_{-r}&=\sum_{\substack{s\ne t\\s,t\ne r}}w_sw_t\ind[a_s=a_t].
\end{align}
The influence return and leave-one-out baseline are
\begin{align}
I_r&=2\ind[a_r=y]-\frac{2}{M-1}\sum_{s\ne r}w_s\ind[a_s=a_r],\\
b_r&=\frac{2C_{-r}}{M-1}-\frac{2P_{-r}}{(M-1)(M-2)}.
\end{align}
With importance weights and advantages detached, the training loss is
\begin{equation}
\mathcal{L}_{\mathrm{RLCD}}=-\frac{1}{M}\sum_{r=1}^{M}
\sg\!\left[w_r(I_r-b_r)\right]\log\pi(a_r).
\label{eq:rlcd}
\end{equation}
Label agreement and the pairwise term determine the update, with each baseline excluding $a_r$. All 32 actions share one forward distribution.

\subsection{Optimization and head refitting}
The English training and validation partitions contain 58,594 and 6,460 clips from Smart Turn v3.1 \citep{smartdata}. Both input orders initialize fresh LoRA parameters and scalar heads on Qwen3-ASR-0.6B. LoRA rank/scaling/dropout is 16/32/0.05 on attention q/k/v/o projections. Stage I uses 4,919 AdamW updates with learning rate $2\times10^{-4}$, 147 warmup updates, cosine decay, weight decay 0.01, and gradient clipping at 1. Batch size is 64, using microbatches of four and 16 accumulation steps.

Stage II uses 1,000 updates at $2\times10^{-5}$ with 30 warmup updates. The main training seed is 1234. For the depth study, each prefix head receives 4,919 matched refitting updates. Results average three head seeds within each backbone seed (1234/2345/3456), followed by an average across the three backbones.

\section{Evaluation Protocol}
\label{sec:evaluation}
\subsection{Benchmark and metrics}
EOT-bench evaluates endpoint decisions at pauses in human-to-agent conversations across 14 languages \citep{eotbench}. Completed turns (EOT) are positive and intermediate pauses (HOLD) are negative. HOLD durations are 0.2--5.0 s, and classification scores are evaluated 200 ms after pause onset. AUC measures score ranking with ties contributing one half; average precision (AP) summarizes precision across EOT recall. Both are reported as percentages.

The policy scan varies 101 confidence thresholds, nine minimum action delays (0.2--1.0 s), and six forced-silence timeouts (1.0--3.5 s), plus 14 silence-only policies, totaling 5,468 candidates. Cut@300ms/600ms minimizes HOLD interruption under a mean EOT waiting budget; lat@5\%cut/10\%cut minimizes waiting under a cutoff-rate budget. Each operating point selects its own policy. Reported latencies measure policy waiting time and exclude model computation.

Endpoint adaptation uses English labels only, and public benchmark validation informed development. Multilingual metrics weight languages equally and use per-language policy selection. These results therefore describe benchmark operating points rather than one global deployment threshold. The 200 ms classification timestamp is distinct from the model's 320 ms online update interval.

\subsection{Comparators and numerical settings}
We compare FD-VAD \citep{fdvad}, DuplexJev-Para/LiveTurn \citep{duplexjev}, and the systems in Table~\ref{tab:main}. A dagger marks published benchmark results; other rows are local evaluations. Deepgram Flux has English results only. Models may differ in context length, training history, and inference interface.

DuplexJev uses native audio-only input with up to 8 s context and the score $P(\mathrm{finished})+P(\mathrm{backchannel})$. Chinese audio uses the native Chinese question; other audio uses the English question. Local inference uses FP32 parameters, BF16 computation, FP32 probabilities, batch size four, and disabled TF32. Input-order and depth comparisons share an evaluation device; separate complete-coverage runs provide system comparisons.

\FloatBarrier

\section{Endpointing Results}
\label{sec:results}
\subsection{Full-model comparison}
Table~\ref{tab:main} reports Chinese and English endpointing results side by side.

\FloatBarrier
\begin{table}[H]
\centering
\caption{Full-model endpointing results for Chinese and English. AUC, AP, and cutoff rates are percentages; latencies are milliseconds. $\dagger$ denotes published results; other rows are local evaluations. --- denotes unavailable results. Bold and underlining mark best and second-best values within each language group. Chinese and English share a system column. Cut300/600 and lat5\%/10\% denote the corresponding budgeted operating points.}
\label{tab:main}
{\small\setlength{\tabcolsep}{1.5pt}
\begin{tabular}{l*{12}{r}}
\toprule
\multirow{2}{*}{System} & \multicolumn{6}{c}{Chinese} & \multicolumn{6}{c}{English}\\
\cmidrule(lr){2-7}\cmidrule(lr){8-13}
 & AUC$\uparrow$ & AP$\uparrow$ & \shortstack{cut$\downarrow$\\300} & \shortstack{cut$\downarrow$\\600} & \shortstack{lat$\downarrow$\\5\%} & \shortstack{lat$\downarrow$\\10\%} & AUC$\uparrow$ & AP$\uparrow$ & \shortstack{cut$\downarrow$\\300} & \shortstack{cut$\downarrow$\\600} & \shortstack{lat$\downarrow$\\5\%} & \shortstack{lat$\downarrow$\\10\%}\\
\midrule
\textbf{JEV-VAD} & \underline{89.4} & \underline{76.7} & \textbf{19.57} & \underline{9.05} & \textbf{772} & \textbf{557} & \underline{94.5} & \underline{89.1} & 16.31 & \underline{9.50} & \underline{847} & \underline{548} \\
DuplexJev-Para & 75.9 & 57.9 & 39.14 & 13.49 & 968 & 678 & 89.7 & 80.9 & 25.53 & 13.05 & 976 & 702 \\
DuplexJev-LiveTurn & 74.9 & 56.6 & 39.14 & 13.16 & 902 & 685 & 89.5 & 80.7 & 25.39 & 12.62 & 960 & 690 \\
LiveKit TD v1$^{\dagger}$ & \textbf{89.9} & \textbf{80.3} & \underline{20.56} & \textbf{8.72} & \underline{799} & \underline{577} & \textbf{96.9} & \textbf{94.1} & \textbf{9.93} & \textbf{4.54} & \textbf{543} & \textbf{295} \\
LiveKit TD v1-mini$^{\dagger}$ & 82.4 & 67.5 & 27.47 & 13.16 & 919 & 691 & 89.0 & 81.3 & 27.80 & 12.06 & 1070 & 698 \\
FD-VAD & 84.5 & 68.3 & 28.62 & 12.17 & 919 & 665 & 90.3 & 81.2 & 24.82 & 13.48 & 1102 & 723 \\
ultraVAD$^{\dagger}$ & 77.6 & 59.1 & 38.82 & 12.83 & 926 & 688 & 88.4 & 79.7 & 27.66 & 11.91 & 899 & 663 \\
SmartTurn v3.2 & 74.5 & 58.2 & 42.93 & 13.16 & 959 & 679 & 84.4 & 73.8 & 35.18 & 14.75 & 1051 & 739 \\
SoulX-Duplug & 67.0 & 52.7 & 48.68 & 15.46 & 1006 & 737 & 70.9 & 57.8 & 51.91 & 20.57 & 1438 & 872 \\
TurnSense v1.1 & 67.7 & 47.7 & 49.84 & 14.97 & 1019 & 740 & 68.6 & 53.4 & 52.91 & 20.43 & 1340 & 902 \\
Silence-only VAD$^{\dagger}$ & --- & --- & 51.15 & 16.28 & 1100 & 800 & --- & --- & 55.60 & 21.70 & 1600 & 1000 \\
Deepgram Flux$^{\dagger}$ & --- & --- & --- & --- & --- & --- & 94.1 & 87.8 & \underline{12.91} & 9.93 & 1151 & \underline{548} \\
\bottomrule\end{tabular}
}

\end{table}

In Chinese, \method obtains cut@300ms of 19.57\%, the lowest value among the compared systems. DuplexJev-Para and DuplexJev-LiveTurn both obtain 39.14\%, while FD-VAD obtains 28.62\%. At a 5\% cutoff budget, \method waits 772 ms, compared with 968 ms for DuplexJev-Para, 902 ms for DuplexJev-LiveTurn, and 919 ms for FD-VAD.

This advantage is specific to the evaluated operating points. LiveKit TD v1 has higher Chinese AUC/AP (89.9\%/80.3\% versus 89.4\%/76.7\%) and a lower cut@600ms (8.72\% versus 9.05\%). Thus, the lowest Chinese cut@300ms does not imply superiority on every discrimination metric or policy budget.

In English, \method reaches cut@300ms of 16.31\%, below DuplexJev-Para (25.53\%) and FD-VAD (24.82\%). However, LiveKit TD v1 reaches 9.93\%, and Deepgram Flux reaches 12.91\%. 

\Needspace{32\baselineskip}
\subsection{Multilingual transfer}
Table~\ref{tab:macro} summarizes full-model transfer across the 14 evaluation languages using equal language weights.

\begin{table}[H]
\centering
\caption{Full-model macro-averaged results over 14 languages. AUC, AP, and cut are in \%; lat is in ms. Cut300/600 and lat5\%/10\% denote the budgeted operating points defined in Section~\ref{sec:evaluation}. The dagger and emphasis conventions follow Table~\ref{tab:main}.}
\label{tab:macro}
{\small\setlength{\tabcolsep}{3pt}
\begin{tabular}{lrrrrrr}
\toprule
System & AUC$\uparrow$ & AP$\uparrow$ & \shortstack{cut$\downarrow$\\300} & \shortstack{cut$\downarrow$\\600} & \shortstack{lat$\downarrow$\\5\%} & \shortstack{lat$\downarrow$\\10\%}\\
\midrule
\textbf{JEV-VAD} & \underline{91.1} & \underline{84.2} & \underline{19.09} & \underline{8.39} & \underline{775.9} & \underline{528.2} \\
DuplexJev-Para & 82.8 & 72.0 & 32.75 & 11.84 & 841.2 & 640.0 \\
DuplexJev-LiveTurn & 82.5 & 71.9 & 33.20 & 11.88 & 840.7 & 642.1 \\
LiveKit TD v1$^{\dagger}$ & \textbf{93.9} & \textbf{89.7} & \textbf{13.99} & \textbf{5.61} & \textbf{624.5} & \textbf{397.7} \\
FD-VAD & 85.3 & 75.2 & 27.63 & 11.82 & 890.1 & 656.7 \\
LiveKit TD v1-mini$^{\dagger}$ & 84.6 & 76.1 & 28.45 & 11.53 & 881.3 & 633.5 \\
ultraVAD$^{\dagger}$ & 79.6 & 70.5 & 37.42 & 13.13 & 851.1 & 659.3 \\
SmartTurn v3.2 & 78.3 & 68.2 & 39.72 & 13.30 & 894.3 & 679.9 \\
TurnSense v1.1 & 58.9 & 50.4 & 50.88 & 17.00 & 1002.6 & 756.1 \\
SoulX-Duplug & 52.8 & 44.5 & 54.37 & 18.16 & 1038.7 & 778.7 \\
Silence-only VAD$^{\dagger}$ & --- & --- & 54.81 & 18.34 & 1085.7 & 821.4 \\
\bottomrule\end{tabular}}

\end{table}
Across 14 languages, \method obtains mean cut@300ms of 19.09\%, compared with 32.75\% for DuplexJev-Para, 33.20\% for DuplexJev-LiveTurn, and 27.63\% for FD-VAD. Its mean lat@5\%cut is 775.9 ms, compared with 841.2 ms for DuplexJev-Para and 890.1 ms for FD-VAD. These results show transfer of English endpoint adaptation beyond the training language.

\begin{figure}[H]
\centering
\includegraphics[width=\linewidth]{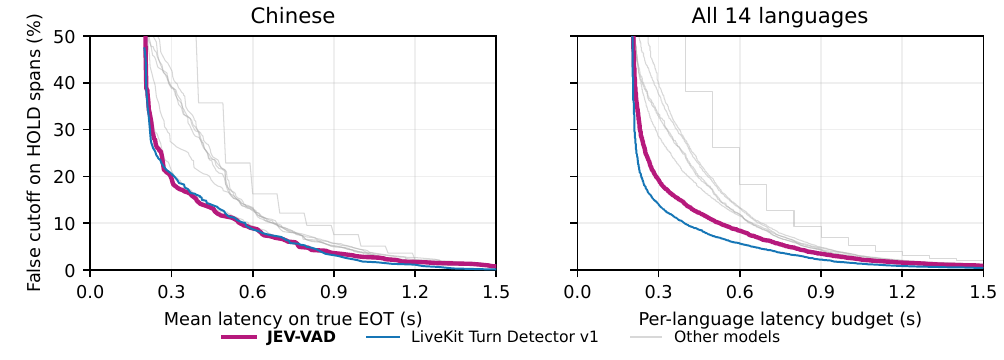}
\caption{Latency--cutoff trade-offs for Chinese (left) and all 14 languages (right). Thick curves denote \method. The multilingual horizontal axis is the per-language latency budget; it is not a measured inference-time axis. Colored thin curves identify comparators with lower cutoff rates at 300 or 600 ms in the multilingual panel, and other comparators are gray.}
\label{fig:pareto}
\end{figure}

LiveKit TD v1 remains stronger on all four reported multilingual operating points, including cut@300ms of 13.99\% and lat@5\%cut of 624.5 ms. Figure~\ref{fig:pareto} shows the policy trade-offs on Chinese and the multilingual aggregate. 

\Needspace{26\baselineskip}
\subsection{Effect of input order}
Table~\ref{tab:order} compares task-first and audio-first conditioning after both adaptation stages.

\FloatBarrier
\begin{table}[H]
\centering
\caption{Paired input-order comparison after SFT and RLCD. AUC/AP/cut are in \%; lat is in ms. The task-first order is better on every listed metric in both languages.}
\label{tab:order}
{\small\setlength{\tabcolsep}{2.4pt}
\begin{tabular}{llrrrrrr}
\toprule
Language & Order & AUC$\uparrow$ & AP$\uparrow$ & \shortstack{cut@\\300ms$\downarrow$} & \shortstack{cut@\\600ms$\downarrow$} & \shortstack{lat@\\5\%cut$\downarrow$} & \shortstack{lat@\\10\%cut$\downarrow$}\\
\midrule
Chinese & $\boldsymbol{q+c+\mathrm{audio}}$ & \textbf{89.4} & \textbf{76.7} & \textbf{19.57} & \textbf{9.05} & \textbf{772} & \textbf{557} \\
Chinese & $\mathrm{audio}+q+c$ & 89.0 & 75.5 & 21.55 & 9.54 & 798 & 587 \\
\midrule
English & $\boldsymbol{q+c+\mathrm{audio}}$ & \textbf{94.5} & \textbf{89.1} & \textbf{16.31} & \textbf{9.50} & \textbf{847} & \textbf{548} \\
English & $\mathrm{audio}+q+c$ & 94.0 & 87.8 & 17.30 & 9.79 & 889 & 585 \\
\bottomrule\end{tabular}}

\end{table}

Task-first conditioning improves all six reported metrics in Chinese and English. Chinese cut@300ms decreases from 21.55\% to 19.57\%, and English cut@300ms decreases from 17.30\% to 16.31\%. Chinese/English AP increases from 75.5\%/87.8\% to 76.7\%/89.1\%.

With readout, token count, and training fixed, these results support placing task instructions before audio. Semantic conditioning and positional effects are not separately isolated.

\Needspace{34\baselineskip}
\subsection{Where does endpoint discrimination become accessible?}
Table~\ref{tab:depth} compares retained decoder prefixes under matched head refitting.

\FloatBarrier
\begin{table}[H]
\centering
\caption{Decoder-prefix results with a fresh head refitted at every depth, including L28. Each value averages three head seeds within each of three backbone seeds, then averages backbones. AUC/AP/cut are in \%; lat is in ms. Chinese and English are shown side by side. Cut300/600 and lat5\%/10\% denote the budgeted operating points. L28 is a refitted-head control, distinct from the full-model system row.}
\label{tab:depth}
{\small\setlength{\tabcolsep}{2.4pt}
\begin{tabular}{l*{12}{r}}
\toprule
\multirow{2}{*}{Depth} & \multicolumn{6}{c}{Chinese} & \multicolumn{6}{c}{English}\\
\cmidrule(lr){2-7}\cmidrule(lr){8-13}
 & AUC$\uparrow$ & AP$\uparrow$ & \shortstack{cut$\downarrow$\\300} & \shortstack{cut$\downarrow$\\600} & \shortstack{lat$\downarrow$\\5\%} & \shortstack{lat$\downarrow$\\10\%} & AUC$\uparrow$ & AP$\uparrow$ & \shortstack{cut$\downarrow$\\300} & \shortstack{cut$\downarrow$\\600} & \shortstack{lat$\downarrow$\\5\%} & \shortstack{lat$\downarrow$\\10\%}\\
\midrule
L00 & 50.0 & 36.1 & 51.15 & 16.28 & 1500 & 800 & 50.0 & 36.2 & 55.60 & 21.70 & 2000 & 1000 \\
L01 & 86.9 & 73.6 & 25.13 & 10.22 & 889 & 613 & 90.8 & 81.2 & 22.90 & 12.26 & 1129 & 713 \\
L02 & 86.9 & 72.9 & 24.84 & 10.34 & 891 & 613 & 93.0 & 86.3 & 20.22 & 10.53 & 997 & 636 \\
L03 & 88.1 & 75.7 & 23.72 & 10.16 & 872 & 607 & 93.1 & 85.7 & 19.54 & 10.28 & 963 & 622 \\
L04 & 87.3 & 73.1 & 22.83 & 9.85 & 883 & 595 & 92.9 & 85.7 & 19.84 & 11.06 & 966 & 648 \\
L05 & 88.3 & 73.9 & 21.40 & 9.96 & 866 & 598 & 92.5 & 84.9 & 21.32 & 11.73 & 968 & 674 \\
L06 & 88.9 & 75.0 & \underline{21.14} & 9.69 & 859 & 586 & 92.9 & 85.5 & 20.47 & 11.47 & 963 & 659 \\
\textbf{L07} & \underline{89.0} & \underline{75.8} & \textbf{21.09} & 9.63 & 863 & 596 & 93.1 & 85.8 & 19.42 & 10.76 & 958 & 644 \\
L14 & 88.8 & 75.7 & 21.38 & \underline{9.56} & \underline{820} & \underline{585} & 94.0 & 87.8 & 18.05 & 9.55 & 899 & 569 \\
L21 & 88.9 & 75.7 & 21.18 & 9.74 & 821 & 591 & \underline{94.3} & \underline{88.4} & 17.62 & \textbf{9.31} & 882 & \textbf{555} \\
\textbf{L28} & \textbf{89.1} & \textbf{76.4} & 21.60 & \textbf{9.36} & \textbf{812} & \textbf{577} & \textbf{94.4} & \textbf{88.8} & \textbf{17.46} & \underline{9.35} & \textbf{869} & \underline{561} \\
\bottomrule\end{tabular}
}

\end{table}

L00 gives AUC of 50.0\% in both languages, while L01 reaches 86.9\% in Chinese and 90.8\% in English. This sharp change is consistent with the no-attention control: the scalar readout requires a decoder attention pathway to access audio. With one block, a refitted head already recovers substantial endpoint discrimination.

Deeper prefixes provide additional gains, especially in English. English AUC rises to 94.4\% at L28, while the best observed Chinese AUC is 89.1\% at L28. Policy metrics are not strictly monotone in retained depth. For example, Chinese cut@300ms is 21.09\% at L07 and 21.60\% at L28, whereas English cut@300ms is 19.42\% at L07 and 17.46\% at L28. These are descriptive means, not significance estimates.

L07 illustrates a selectable quality--resource budget rather than a lossless replacement. Relative to L28, English cut@300ms worsens by 1.96 percentage points and lat@5\%cut rises from 869 to 958 ms. Chinese cut@300ms improves by 0.51 points, but Chinese lat@5\%cut rises from 812 to 863 ms. Retaining seven blocks therefore preserves one Chinese interruption operating point while changing the waiting-time trade-off. Its suitability depends on the deployment language and the chosen interruption budget.

\FloatBarrier
\Needspace{34\baselineskip}
\section{Parameter and Computation Budgets}
\label{sec:efficiency}
Table~\ref{tab:cost} reports unique deployed parameters and matrix-operation counts for the stated input lengths.

\begin{table}[H]
\centering
\caption{Parameters and single-forward matrix computation for 2.56 s audio. Decoder computation excludes the encoder; output computation is separate. DuplexJev uses its Chinese prompt. M denotes millions of parameters.}
\label{tab:cost}
\begin{tabular}{lrrr}
\toprule
Model / prefix & \shortstack{Params\\(M)} & \shortstack{Decoder\\GFLOPs} & \shortstack{Output\\FLOPs}\\
\midrule
DuplexJev-Para & 4219.71 & 747.333 & $7.78\times10^{8}$ \\
DuplexJev-LT & 4219.71 & 747.333 & $7.78\times10^{8}$ \\
L00 & 341.96 & 0.000 & 2048 \\
L01 & 357.86 & 2.364 & 2048 \\
L02 & 373.75 & 4.728 & 2048 \\
L03 & 389.65 & 7.092 & 2048 \\
L04 & 405.54 & 9.456 & 2048 \\
L05 & 421.43 & 11.820 & 2048 \\
L06 & 437.33 & 14.184 & 2048 \\
L07 & 453.22 & 16.548 & 2048 \\
L14 & 564.49 & 33.095 & 2048 \\
L21 & 675.75 & 49.643 & 2048 \\
L28 & 787.01 & 66.191 & 2048 \\
\bottomrule\end{tabular}

\end{table}

L07 contains 453.22M parameters versus 787.01M for L28, a 42.41\% reduction. Total compression is smaller than the fraction of removed decoder blocks because the encoder, projector, and required embeddings remain present. With 73 input tokens, decoder matrix computation decreases from 66.191 to 16.548 GFLOPs, a 75\% reduction. The scalar head requires 2,048 FLOPs, compared with $7.78\times10^8$ for DuplexJev's vocabulary projection. DuplexJev's decoder count uses 102 tokens and a different architecture.

Counts use two FLOPs per multiply--accumulate and include linear/LoRA operations and dense QK/AV products. Encoder computation, preprocessing, normalization, activations, softmax, and scalar output postprocessing are excluded. The scalar head removes vocabulary-sized output work, while prefix pruning removes decoder blocks. These resource reductions do not establish wall-clock speedup; end-to-end inference time remains to be measured.

\section{Discussion and Limitations}
\label{sec:discussion}
The depth study measures endpoint information accessible after matched head refitting. It does not establish that the original final-layer head can be transferred to any depth without adaptation. The language-dependent policy trade-offs also preclude treating one retained depth as universally optimal. Depth results are bilingual means without dispersion estimates; the 14-language comparison uses the full model.

All variants inherit SFT--RLCD training. An SFT-only ablation and calibration-specific measurements are needed to isolate RLCD's contribution. Public validation informed development, and policies are selected separately by language and budget; independent held-out evaluation would strengthen deployment claims. The system comparison includes clearly marked published references alongside local runs, and reported policy delays exclude inference time.

\Needspace{11\baselineskip}
\section{Conclusion}
\method combines task-first conditioning, a scalar decision head, and decoder-prefix pruning for efficient semantic endpointing. The full model, adapted with English endpoint labels, achieves Chinese cut@300ms of 19.57\% and a 14-language mean of 19.09\%. A paired input-order comparison favors placing task instructions before audio. Matched multi-seed prefix experiments locate useful endpoint discrimination in early decoder blocks and quantify the trade-off associated with seven-block retention: 42.41\% fewer total parameters and 75\% less decoder matrix computation, with language- and budget-dependent changes in endpoint quality. These findings support selecting a decoder budget for the endpointing task rather than inheriting the full generation pathway by default.

\FloatBarrier
\bibliographystyle{plainnat}
\bibliography{references}
\end{document}